\documentclass[
superscriptaddress,
 preprint,
 amsmath,amssymb,
 aps,
]{revtex4-2}

\usepackage{graphicx} 
\usepackage[utf8]{inputenc}
\usepackage{amsmath}
\usepackage{dcolumn}
\usepackage{bm}
\usepackage{siunitx}
\usepackage{cancel}
\usepackage[normalem]{ulem}
\usepackage{footnote}
\usepackage{float}
\usepackage{xcolor}
\usepackage{hyperref}

\hypersetup{
    colorlinks = true,
    linkcolor= blue,
    citecolor= blue,
    urlcolor= blue
}

\usepackage{cleveref}
\usepackage{natbib}
\usepackage{booktabs}
\usepackage{bbding}

\begin{document}

\title{Multicell-Fold: geometric learning in folding multicellular life}

\author{Haiqian Yang}
\affiliation{Department of Mechanical Engineering, Massachusetts Institute of Technology, 77 Massachusetts Ave., Cambridge, MA 02139, USA}

\author{Anh Q. Nguyen}
\affiliation{Department of Physics, Northeastern University, Boston, MA 02115, USA}

\author{Dapeng Bi}
\affiliation{Department of Physics, Northeastern University, Boston, MA 02115, USA}

\author{Markus J. Buehler}
\affiliation{Department of Mechanical Engineering, Massachusetts Institute of Technology, 77 Massachusetts Ave., Cambridge, MA 02139, USA}
\affiliation{Laboratory for Atomistic and Molecular Mechanics (LAMM), Massachusetts Institute of Technology, 77 Massachusetts Ave., Cambridge, MA 02139, USA}
\affiliation{Center for Computational Science and Engineering, Schwarzman College of Computing, Massachusetts Institute of Technology, 77 Massachusetts Ave., Cambridge, MA 02139, USA}

\author{Ming Guo}\thanks{guom@mit.edu}
\affiliation{Department of Mechanical Engineering, Massachusetts Institute of Technology, 77 Massachusetts Ave., Cambridge, MA 02139, USA}

\date{\today}

\begin{abstract}

During developmental processes such as embryogenesis, how a group of cells fold into specific structures, is a central question in biology. However, it remains a major challenge to understand and predict the behavior of every cell within the living tissue over time during such intricate processes. Here we present a geometric deep-learning model that can accurately capture the highly convoluted interactions among cells. We demonstrate that multicellular data can be represented with both granular and foam-like physical pictures through a unified graph data structure, considering both cellular interactions and cell junction networks. Using this model, we achieve interpretable 4-D morphological sequence alignment, and predicting cell rearrangements before they occur at single-cell resolution. Furthermore, using neural activation map and ablation studies, we demonstrate cell geometries and cell junction networks together regulate morphogenesis at single-cell precision. This approach offers a pathway toward a unified dynamic atlas for a variety of developmental processes.

\end{abstract}

\maketitle
\clearpage

\section*{Main}

How living tissues such as embryos spontaneously self-organize into distinct yet robust structures is the central question of developmental biology and one of the greatest unsolved problems~\cite{keller2002shaping,zhu2020principles,liu2024morphogenesis,gilmour2017morphogen,keller2012physical,trepat2018mesoscale}. The various forms of multicellular species emerge from the spatiotemporal dynamic interactions among tens, hundreds, thousands of, or more cells~\cite{stern2022deconstructing,mongera2018fluid,zenker2018expanding,lim2020keratins,rozbicki2015myosin}. In contrast to folding protein structures from a chain of amino acids where modern deep-learning models have achieved accuracy at levels of single atoms~\cite{abramson2024accurate,jumper2021highly}, it has not yet been possible to predict the self-organization of multicellular structures from a population of cells with accuracy at the scale of single cells.

One particularly important local mechanism that sculpts living tissues is cell rearrangements, meaning two adjacent cells lose their shared junction and become non-adjacent; massive local cell rearrangements are responsible for rapid tissue elongation during embryogenesis in almost all multicellular species~\citep{campas2024adherens,bi2015density,mongera2018fluid,kim2021embryonic,mongera2023mechanics,bertet2004myosin,keller2002shaping,blanchard2009tissue,rozbicki2015myosin,rauzi2008nature,mongera2018fluid,rozbicki2015myosin,wang2020anisotropy,etournay2015interplay,herrera2023tissue,cupo2024signatures}. A long-standing hypothesis is that the geometries and dynamics of cells and their junctions may contain critical information that determines cell rearrangements~\cite{stern2022deconstructing,walck2014cell,campas2024adherens,brauns2024geometric,cupo2024signatures}. However, generating such future prediction at single-cell resolution is challenging both for experimental studies and mechanistic theories, which have remained elusive especially due to the active and out-of-equilibrium nature of living systems~\cite{trepat2018mesoscale,campas2024adherens,bi2015density,wang2020anisotropy,mongera2018fluid,kim2021embryonic,campas2014quantifying}. Therefore, it has not yet been possible to predict where and when local cell rearrangements happen at single-cell resolution prior to their occurrence. Being able to predict this will provide crucial insights into understanding morphogenesis, the ``folding" of living tissues during development and disease progression.

To build a foundational data-driven model to decipher multicellular folding processes, a standardized data structure is critical, but has remained elusive. Historically, researchers have developed mechanistic frameworks that originate from two separate physical pictures, i.e. granules and foams. From the granular perspective, a multicellular system is viewed as a point cloud of cells, where glassy dynamics and packing behaviors are extensively studied~\cite{angelini2011glass,atia2018geometric,brandstatter2023curvature,tang2024nuclear}. A granular perspective has also been shown to be useful when examining large areas of histological tissue samples~\cite{hu2024unsupervised}. In contrast, in the foam-like picture, a multicellular system is viewed as a graph of actual cell edges connecting through vertices~\cite{lecuit2007cell,kim2021embryonic,rozman2020collective,wang2024cadherin}; dating back to the early days when D'Arcy Thompson examined morphogenesis~\cite{thomson1917growth}, an important aspect of this notion is surface tension which continues to be appreciated in the embryo research today~\cite{firmin2024mechanics,brauns2024geometric,martin2009pulsed,yevick2019structural}. Despite the groundbreaking studies from both angles highlighting the physical aspect of multicellular folding processes, it is unclear how to unify these two aspects within one framework.

Here we present a foundational geometric deep-learning workflow tailored to predict mesoscale multicellular evolution, where both the granular and foam-like physical pictures can be represented within a unified dual-graph data structure (Fig.~\ref{fig:overview}). With this modeling approach, for a time-lapsed \textit{Drosophila} 3-D whole-embryo dataset, we first demonstrate unsupervised and interpretable geometric video sequence alignment (Fig.~\ref{fig:flyAlign}), which enables large-scale 4-D morphological sequencing and data registration. Furthermore, a model is trained on one 3-D \textit{Drosophila} whole-embryo video with labeled cell rearrangements, and then validated on a different \textit{Drosophila} whole-embryo video to identify future rearrangements that have not yet happened, and we demonstrate an unprecedented ability to predict cell rearrangements at single-cell resolution before they occur, with an overall accuracy $>82\%$ (Fig.~\ref{fig:flyT1}, {\color{blue}Supplementary Materials} Table~S1 and Table~S2). Our workflow is the first algorithm, to our knowledge, that can achieve close-to-experiment accuracy in predicting multicellular folding processes at single-cell resolution. Beyond prediction, using an analysis of neural activation maps and ablation studies, we also identify key geometric signatures underlying these crucial developmental processes.

\subsection*{A geometric deep-learning workflow}

\begin{figure}[h!]
    \centering
    \includegraphics[width=.98\textwidth]{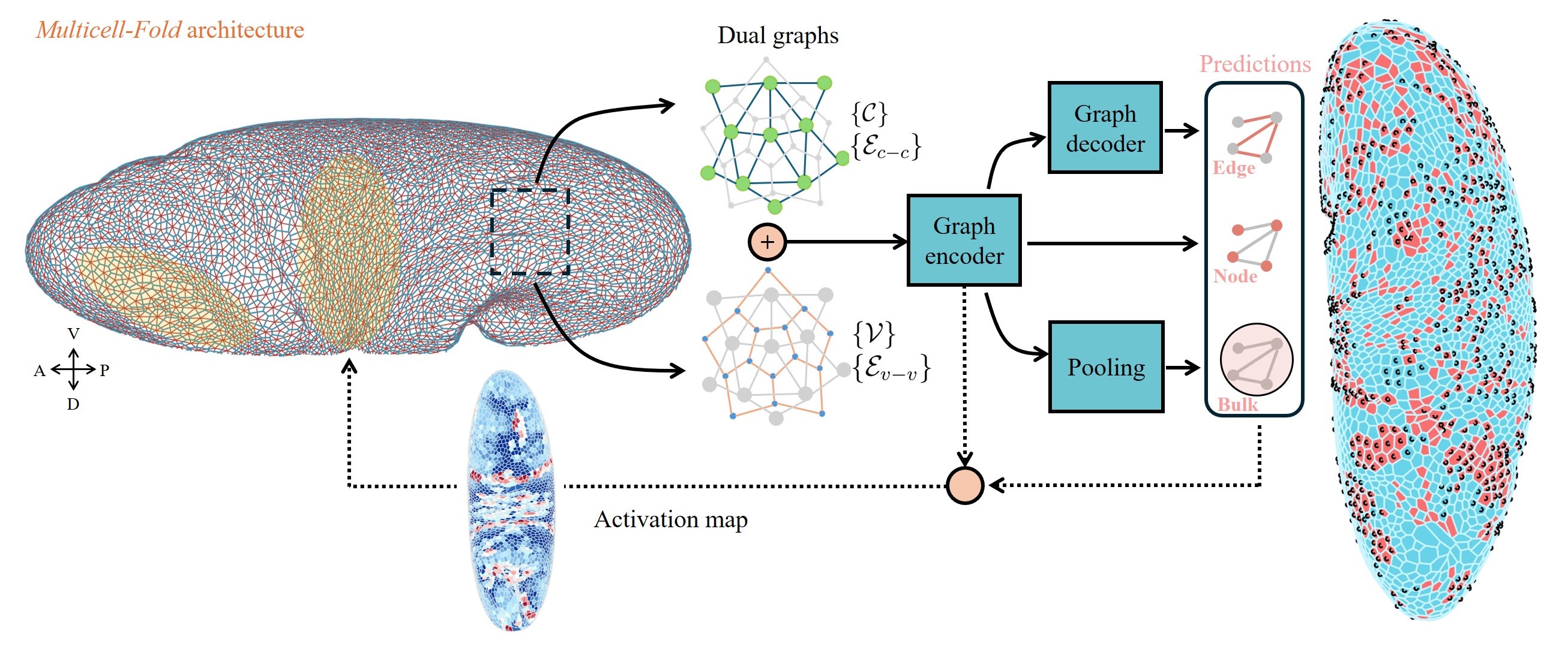}
    \caption{Overview of our multicellular folding algorithm (Multicell-Fold). (From left to right) A representative snapshot of a developing embryo (\textit{Drosophila}, imaging and tracking data from~\cite{stern2022deconstructing}). To build a data-driven model to study these mesoscale living systems, we propose that these data can be represented as a dual graph structure, consisting of a primary graph of cells $\{\mathcal{C}\}$ and cell-cell adjacency $\{\mathcal{E}_{c-c}\}$, and an auxiliary graph of vertices $\{\mathcal{V}\}$ and cell edges $\{\mathcal{E}_{v-v}\}$. The two graphs are combined as inputs for the graph encoder, whose outputs can be used as node predictions. Alternatively, subsequently using a decoder or pooling operation generates edge or tissue-level prediction. An example prediction is shown, where orange indicates our model prediction of future cell rearrangements, and true cell rearrangement is indicated with black circles. (From right to left) Using a trained model, an activation map can be used to visualize the regional information the model has used to make predictions.}
    \label{fig:overview}
\end{figure}

The granular and foam-like representations of multicellular living matter each have their strength; gene and protein expressions, as well as cell biophysical quantities such as cell migration speed and cell geometries, are naturally associated with each cell, while cell junction proteins, geometries, and tensions are more related to cell edges. To unify these two aspects, we define a dual-graph representation as shown in Fig.~\ref{fig:overview},

\begin{equation}\label{Eq:graphStructure}
\begin{aligned}
    & \text{nodes} \quad \{\mathcal{C}, \mathcal{V}\}_t\\
    & \text{edges} \quad \{\mathcal{E}_{c-c}, \mathcal{E}_{v-v}, \mathcal{E}_{c-v}\}_t\\
    & \text{tissue quantity}\quad \{ \Lambda \}_t
\end{aligned}
\end{equation}
where $\mathcal{C}$ is a list of cells, $\mathcal{V}$ is a list of vertices, $\mathcal{E}_{c-c}$ is a list of cell-cell adjacency, $\mathcal{E}_{v-v}$ is a list of cell edges, and $\mathcal{E}_{c-v}$ is a list of cell-vertex adjacency. $\Lambda$ is a list of tissue-level variables, which, for example, can be a development time for an embryo, tissue-level stress, or a sample type. The subscript $_t$ denotes frame $t$.

Notably, this data structure (Eq.~\ref{Eq:graphStructure}) is particularly well-suited for geometric deep learning. From a data-driven point of view, a majority of questions in mesoscale developmental biology such as embryogenesis can then be characterized as \textit{graph inference}; a learning algorithm is supposed to learn features from historical or partial graphs to make predictions of future or other partial graphs. For example, many tasks of particular biological and physiological interest in these systems can be classified as node, edge or bulk prediction tasks (Table~\ref{tab:tasks}). A combination of graph neural networks (GNN) structures (e.g. various types of graph convolutional layers, graph autoencoder) has been shown to allow for flexible predictions of complex physical and biological systems~\cite{wang2023scientific,kipf2016semi,kipf2016variational,corso2020principal,velivckovic2017graph,shi2020masked,bapst2020unveiling,yang2022linking,hu2024unsupervised,vinas2023hypergraph,baker2023silico}. However, these existing models are typically designed for molecules, proteins, transcriptomics, and knowledge graphs; directly applying these models cannot account for the distinct data structure of mesoscale multicellular living systems such as developing embryos.

\begin{table}[h!]
    \centering
    \begin{tabular}{|c|c|}
    \hline
     Prediction type    &  Example                                     \\
    \hline
     Bulk               &  \textbf{developmental time alignment}       \\
                        &  healthy and diseased classification         \\
                        &  tissue stress                               \\
    \hline
     Node               &  cell position                               \\
                        &  cell fate                                   \\
                        &  cell multi-omics                            \\
    \hline
     Edge               &  \textbf{cell rearrangement}                 \\
                        &  tension inference                           \\
                        &  leader-follower identification              \\
                        &  lineage trace                               \\
     \hline
    \end{tabular}
    \caption{Inference tasks in multicellular systems.}
    \label{tab:tasks}
\end{table}

Physically, the behavior of each cell not only depends on its own identities, but can also depend on the identities of its neighbors, and recursively on its neighbors' neighboring cells, and so on. To resolve such highly convoluted information, in our workflow (Fig.~\ref{fig:overview}), we employ a graph encoder consisting of multiple message-passage layers (e.g. multi-headed graph transformer layers) as the backbone of our model, which aggregates highly heterogeneous information in space into high dimensional hidden states on each cell. The behavior of a pair of cells (e.g. rearrangement) could be inferred from whether the two cells have matching hidden states, in which case we employ a graph decoder for predicting the behavior of each pair of neighboring cells. Besides, tissue-level states can be viewed as a property emerging from the high-dimensional cell identities among all the cells in the tissue, in which case we employ the pooling operation and a multi-layer perceptron to aggregate hidden cell states into tissue-level labels. With this workflow, in the following, we show automatic and interpretable video sequence alignment as an example of tissue-level tasks; we further show predicting local cell rearrangement as an example of edge prediction.

\subsection*{Interpretable unsupervised geometrical video sequence alignment}

\begin{figure}[h!]
    \centering
    \includegraphics[width=.98\textwidth]{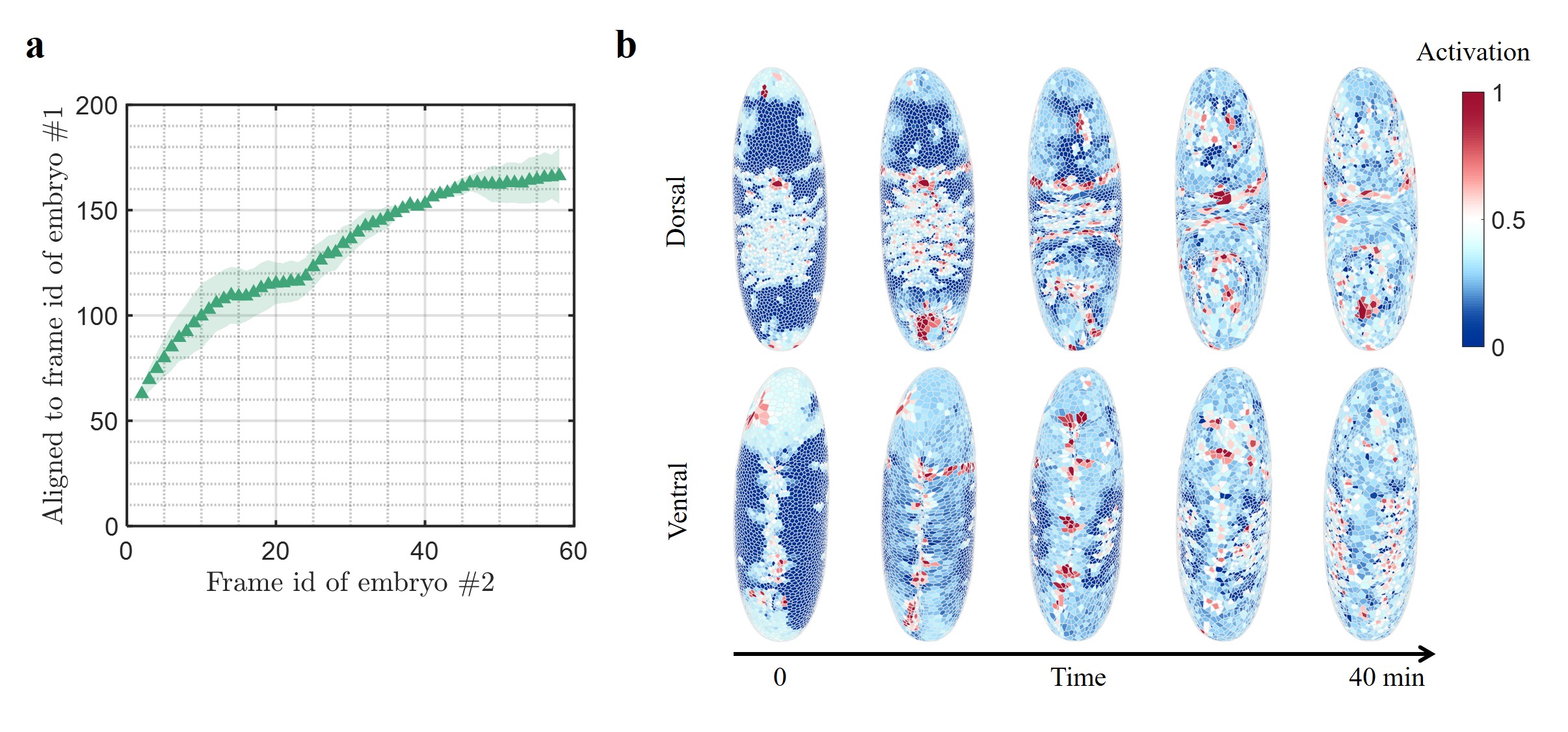}
    \caption{Interpretable geometric video sequence alignment using activation map. (a) Align the two embryo sequences. Three independent models are trained; the markers indicate the mean values and the dashed color indicates the standard deviation. (b) The activation map visualizes the features the model used to make the alignment prediction.}
    \label{fig:flyAlign}
\end{figure}

During embryogenesis, global-level predictions have broad applications, including classification tasks such as distinguishing healthy and diseased samples, regression tasks such as defining developmental time, and inferring tissue-level mechanical or dynamical features. In particular, automatic and high-throughput multiple sequence alignment is critical for ultimately creating a large multicellular folding model. Existing alignment methods rely on computing similarities in the flow fields~\cite{mitchell2022morphodynamic}, and some recent development applies convolutional neural nets on 2-D brightfield images~\cite{toulany2023uncovering}, which does not consider the complex 3-D structure, cellular geometries, and both local and long-range cell-cell interactions that are crucial for development. Here we demonstrate the use of our geometric-learning workflow in solving this challenge, and we show how this can be achieved in an automatic, unbiased, unsupervised, and interpretable way.

To demonstrate this, we perform video sequence alignment of two 3-D time-lapsed whole \textit{Drosophila} embryos during gastrulation from a recent experimental study~\cite{stern2022deconstructing}; two embryos are imaged at arbitrary orientations and different frame rates, and they start to evolve at different times ({\color{blue}Supplementary Materials} Fig.~S4). Using our workflow, we train a regressional model on one embryo as a reference, with its own frame ID as labels. Applying the model to each frame of the second embryo outputs the ID of the most aligned frame from the first embryo, which aligns the developmental time of the two embryos (Fig.~\ref{fig:flyAlign}a). To interpret the alignment results, in a trained model, we adapt the class activation method~\cite{zhou2016learning}; we visualize the weighed activation of the last graph encoding layer before the global pooling operation. The activation map shows clearly that the model captures the ventral furrow and dorsal features (Fig.~\ref{fig:flyAlign}b). An impressive fact is that the model has learned to focus on these features even though it is only trained to regress frame ID, which is very little information.

Automatic video sequence alignment for live imaging of developing tissues is important especially when provided with a large amount of data with different imaging settings. For example, transcriptomic and proteomic data are usually created with fixed samples~\cite{tanay2017scaling,rodriques2019slide,staahl2016visualization}, making it difficult to analyze their dynamics. Our workflow provides a pathway to connect live and fixed datasets, harnessing the multicell morphology at single-cell resolution and complex spatial patterns that are not accessible to traditional alignment methods. This alignment will also allow us to combine live and fixed datasets to visualize the entire spatial and temporal evolution of multi-omics, during the development of embryos, tissues, and organs.

\subsection*{Predict local cell rearrangements before they occur at single-cell resolution}

\begin{figure}[h!]
    \centering
    \includegraphics[width=.98\textwidth]{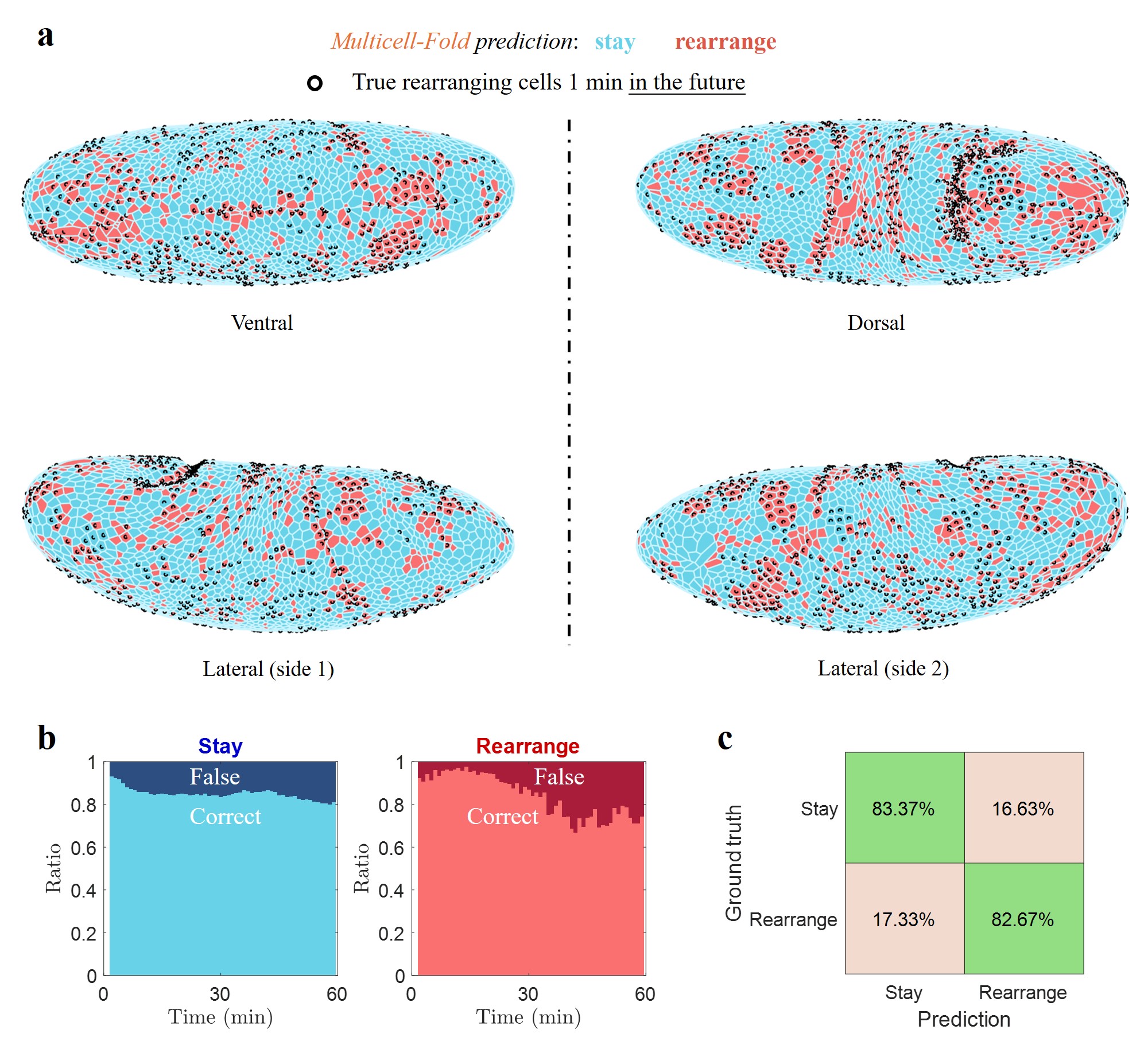}
    \caption{Predicting local cell rearrangement. (a) Example prediction at different angles of a 3-D \textit{Drosophila} embryo. The embryo at $t=30$ min is shown here, and consecutive predictions over time are shown in {\color{blue}Supplementary Video 1}. Cells predicted to lose cell-cell junction with any of their neighbors in the next 1 minute are colored in orange. (b) Model accuracy over time. (c) Average accuracy.}
    \label{fig:flyT1}
\end{figure}
In addition to the global behavior captured by our video sequence alignment, local activities also play a crucial role in tissue dynamics. The importance of these local activities is evident in the establishment of tissue morphology, which is facilitated by massive local cell rearrangements identifiable through the loss of cell-cell junctions. With our versatile and multifunction workflow, our next goal is to pinpoint each single cell-cell junction that will be lost in the near future, through which we also hope to identify interpretable key geometric features underlying these local reorganization events. It is worth noting that while raw coordinates can be directly provided as inputs, it is beneficial to make use of the symmetries of the problem to construct the inputs. Here we constrain input within a subset of invariant quantities ({\color{blue} Materials and Methods}); this will guarantee that the model is invariant of any rigid-body rotations of the whole sample. To identify an efficient invariant data representation that is a subset of Eq.~\ref{Eq:graphStructure}, we use a synthetic dataset of 2-D cell monolayers where we can systematically vary the biophysical parameters and study the model performance with different representations ({\color{blue} Materials and Methods}). Using our workflow, we pinpoint the loss of cell-cell junctions in this simulated dataset ({\color{blue}Supplementary Materials} Fig.~S2). Through a comprehensive ablation study, we observe that the model performance is sensitive to cell area, perimeter, and cell edge length, while it is less affected by cell-cell distance and junction elastic tension, regardless of whether the simulated cell monolayer is fluid-like or solid-like ({\color{blue}Supplementary Materials}, Fig.~S1). We also observe that, while the full dual-graph representation always performs the best, an efficient representation through concatenating actual cell edges on their dual cell-cell adjacency does not significantly harm the model performance ({\color{blue}Supplementary Materials}, Fig.~S1). Nevertheless, the latter reduces the graph size to 1/3 of its original size, which is much more efficient in computation memory and time.

Using this efficient data representation, on the \textit{Drosophila} data we challenge our learning algorithm to predict each single lost cell-cell junction in the future in these embryos~\cite{stern2022deconstructing}. During gastrulation, the \textit{Drosophila} embryos are subjected to a complex combination of internal stresses and rapid deformations, which are accompanied by a large number of local rearrangement events. Pinpointing these local events prior to their occurrence in such a rapid development process has not been possible. On the 3-D video data lasting 60 minutes (1 frame/minute) capturing the whole gastrulation process, from each frame of the video, we seek to pinpoint each pair of cells that will rearrange through losing their shared cell-cell junction in the next 1 minute.

Here we train the model on one of the embryos and test the model on the other. Throughout the entire gastrulation process in the test set from Embryo \#2, a total of 41,~544 cell-cell junctions are lost within a total number of 0.75 million cell-cell junctions (summed across frames, {\color{blue}Supplementary Materials} Table~S1). Even trained on only a single embryo, the model achieves remarkable performance with accuracy $>82\%$ (Fig.~\ref{fig:flyT1}). The video data of each embryo is a giant graph containing enormous nodes and rearrangement events from which the governing rules can be inferred.

Furthermore, we train models to predict loss of cell-cell junction that will occur within 2, 3, 4, and 5 minutes in the future; while the performance decreases with longer time into the future, the model still achieves competitive accuracy $>70\%$ even at 5 minutes in the future ({\color{blue}Supplementary Materials} Fig.~S7). It is worth noting that global coordinates are not directly provided as inputs into the model, therefore, the accurate prediction also indicates that local graph geometries and local cell dynamics may contain important information about these rearrangement events.

\begin{figure}[h!]
    \centering
    \includegraphics[width=.8\textwidth]{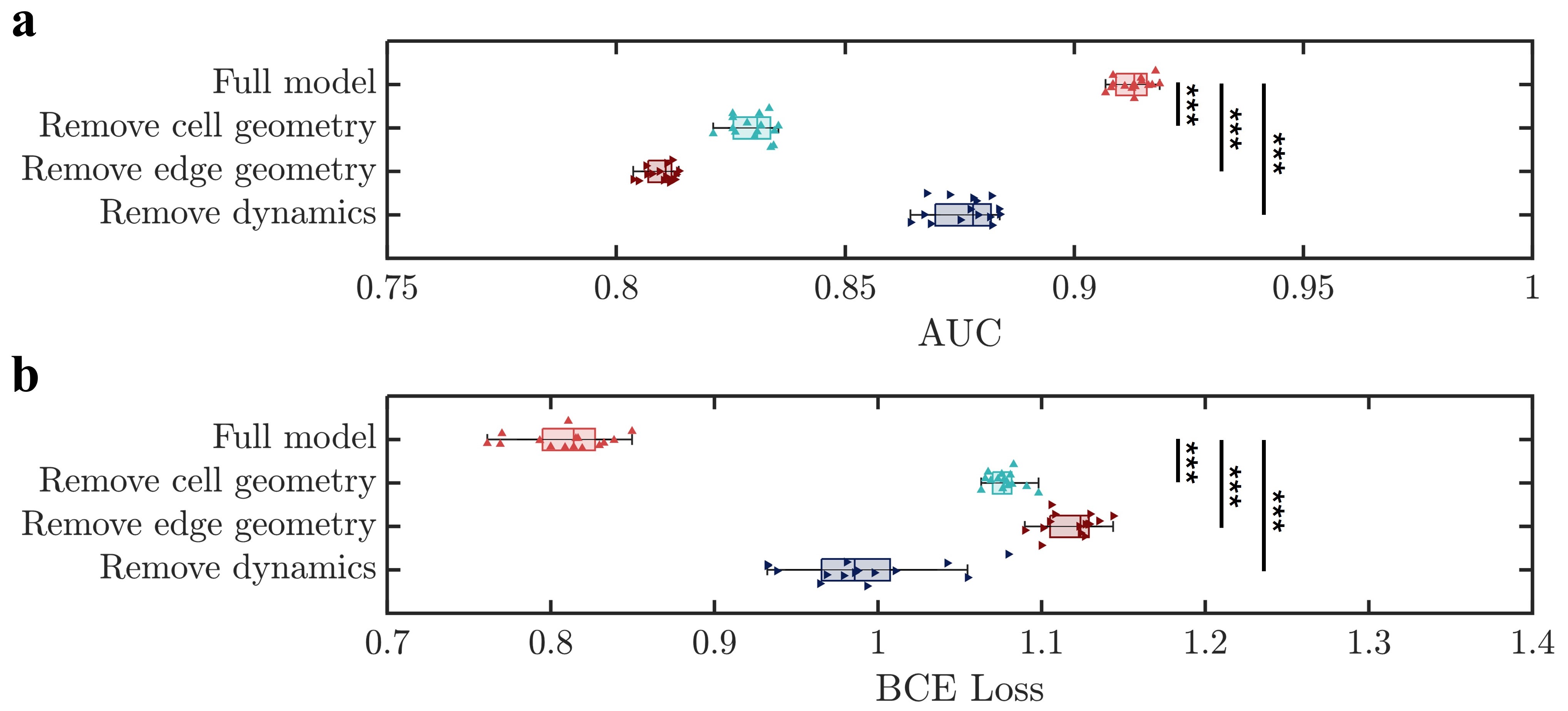}
    \caption{Ablation study. (a) AUC and (b) BCE loss. For each class, we train 3 independent models using 3 random seed numbers for 500 epochs and report the model performance of the last 5 epochs (n=15). One-way ANOVA tests are performed. (***: p-value $<0.001$).}
    \label{fig:mainAblation}
\end{figure}

Three types of information are used in the \textit{Drosophila} tasks, which are cell geometry, cell edge geometry, and dynamic information ({\color{blue} Materials and Methods}). The full model is trained and tested with all the information provided, and it achieves highly accurate prediction with AUC $0.913\pm 0.004$ and a BCE loss $0.809\pm 0.026$ on the test set. To understand the contribution of each input component, we perform an ablation study by training and testing separate models with partial information ablated from both the train and test sets (Fig.~\ref{fig:mainAblation}, see {\color{blue}Materials and Methods} for details of ablation). In the vertex model framework, it has been argued that cell edge length is proportional to the additional stress to yielding and the distribution of the cell edge length is therefore an indicator for tissue instability~\cite{Popovic_2021}. Intuitively, two cells connected through a short cell junction are more likely to rearrange as it requires minimal force to shrink the shared edge. This is indeed reflected in ablating cell edge length from the input, decreasing AUC to $0.810\pm 0.003$ and increasing the BCE loss to $1.118\pm 0.016$. However, we note that solely using edge length as an analytical criterion cannot produce any useful prediction, with an AUC $\sim$ 0.511 which is just slightly better than random guessing, implying that using edge length distribution as the sole predictor for tissue instability is insufficient. This is in contrast to the highly accurate performance of using geometric deep learning, indicating that rearrangement is governed through a convoluted interaction of short edges in space, rather than the information on a single edge. This is consistent with our understanding of the plasticity of amorphous materials, that collective rearrangement is an avalanche-type behavior governed by the complex interplay between spatial patterns of defects~\cite{falk2011deformation} and the stress redistribution \cite{richard2023mechanical}. In addition to the cell edge length, we also expect cell shape to play an important role in tissue dynamics. Recently, a density-independent rigidity transition governed solely by cell shape has been shown~\cite{bi2015density,park2015unjamming}. Here, we find that removing cell geometry also has a prominent influence on the performance, decreasing AUC to $0.830\pm 0.004$, and increasing the BCE loss to $1.077\pm 0.009$, which suggests that rearrangement does not solely depend on the edge network, but also depends on the geometries of the cells connected by the edges. This supports recent theoretical work showing that cell shapes are indicative of tissue fluidity, with more fluid-like packs of cells being more susceptible to rearrangements~\cite{huang2022shear}. Furthermore, we observe that the dynamic information also has a significant influence, and removing it decreases AUC to $0.876\pm 0.007$, and increases the BCE loss to $0.990\pm 0.043$, which concurs with recent studies showing cell dynamics and tissue flows are important signatures during embryogenesis~\cite{lefebvre2023learning,guirao2015unified,etournay2015interplay}. Our workflow harnesses all these different types of information in a highly spatially convoluted way, which can not be achieved by previous methods. 

Notably, these results using experimental data are consistent with our ablation study using the synthetic data ({\color{blue}Supplementary Materials} Fig.~S1). To further validate this observation, we also change the model architecture to using Principal Neighborhood Aggregation (PNA) convolutional layers instead of graph transformer layers on the \textit{Drosophila} dataset. While graph transformer slightly outperforms PNA, among different ablation groups we observe the same trends ({\color{blue}Supplementary Materials} Fig.~S6). In summary, the ablation study suggests that junction edge geometry, cell geometry, and their dynamics are critical features that regulate local cell rearrangements; they provide mutually independent information that cannot be replaced by one another.

\subsection*{Discussions}

We propose a dual-graph modeling strategy as a suitable data structure for mesoscale multicellular living systems such as embryos, taking into account both the granular and foam-like physical pictures of these systems. Based upon this notion, we present a geometric deep-learning approach to decipher the dynamics of morphogenesis at single-cell resolution, that is, \textbf{multicellular folding}. Using this approach, we demonstrate interpretable morphological video sequence alignment, which could provide a foundation for large-scale 4-D morphological sequencing of developing embryos and other tissues. Furthermore, in both 4-D whole \textit{Drosophia} embryogenesis data and synthetic 2-D cell monolayer data, we achieve accurate prediction of local cell rearrangement events in the future at single-cell resolution, with an efficient data representation for fast and efficient learning. Using ablation study on both experimental and synthetic data, we identify cell junction geometry, cell geometry, and their dynamics as important geometric features that underlie the local cell rearrangement events.

Our approach offers an innovative data-driven approach to study mesoscale morphogenesis, by harnessing the multicellular geometric structures in the neighborhood of every single cell to make useful predictions of the future cell behaviors; such an angle is distinct from recent studies to infer continuum equations~\cite{supekar2023learning,lefebvre2023learning}, or individual cell trajectories~\cite{lachance2022learning,bruckner2020inferring,frishman2020learning}. It has remained a challenge to create a single-cell multi-omic atlas for embryogenesis that can be used to infer the behavior of each single cell \textit{in situ} and over time. Our results here suggest that geometric deep learning algorithms may help to achieve this goal. Moreover, our results reveal the possibility of designing a general model for folding multicellular life, as is today with folding proteins; such a model might allow \textit{in silico} drug screening of morphologically-related diseases, and potential \textit{de novo} design of multicellular structures for diverse applications.

\subsection*{Materials and Methods}

A detailed description of the methods is provided in the {\color{blue}Supplementary Materials}. A summary is as follows.

\textit{Efficient invariant representation.}---
The raw graphs contain
\begin{equation}
\begin{aligned}
    &\mathcal{N} = \{\mathcal{C}, \mathcal{V}\} \quad \text{with coordinates} \quad \{\mathbf{x}\}, \\
    &\mathcal{E} = \{\mathcal{E}_{c-c}, \mathcal{E}_{v-v}, \mathcal{E}_{c-v}\},
\end{aligned}
\end{equation}
where $\mathcal{C}$ is a list of cells, $\mathcal{V}$ is a list of vertices, $\mathcal{E}_{c-c}$ is a list of cell-cell adjacency, $\mathcal{E}_{v-v}$ is a list of cell edges, and $\mathcal{E}_{c-v}$ is a list of cell-vertex adjacency.

We consider the invariant quantities as node embeddings and edge attributes, which are cell area $a$ and cell perimeter $p$ on $\{\mathcal{C}\}$, cell-cell distance $l_{c-c}$ on $\{\mathcal{E}_{c-c}\}$, cell edge length $l_{v-v}$ on $\{\mathcal{E}_{v-v}\}$ and cell-vertex distance $l_{c-v}$ on $\{\mathcal{E}_{c-v}\}$.  

We further identify an efficient representation ({\color{blue}Supplementary Materials}), which is
\begin{equation}
\begin{aligned}
    &\mathcal{N} = \{\mathcal{C}\} \quad \text{with node embeddings} \quad \{a, p\}, \\
    &\mathcal{E} = \{\mathcal{E}_{c-c}\} \quad \text{with edge attributes} \quad \{l^*_{v-v}\}, 
\end{aligned}
\end{equation}
where we have assigned cell edge length to its dual cell-cell adjacency, and weighted the edge length $l^*_{v-v}=\exp{(-l_{v-v})}$ to encourage the model to pay more attention to the pattern of the short edges. Unless otherwise noted, we use this efficient invariant representation throughout the paper.

In all the \textit{Drosophila} tasks, the input graphs contain a minimal set of invariant quantities, which are
\begin{equation}
    \text{Node embedding} \quad \{a,\dot{a},p,\dot{p}\} \text{, and edge attribute} \quad \{l^*_{v-v}, \dot{l^*_{v-v}}\} ,
\end{equation}
where the $\dot{}$ denotes the difference of the quantity compared to the previous frame, which is used to inform the model of the system dynamics.

\textit{Ablation study.}---
In the ablation study, removing cell edge geometry refers to removing ($\{l^*_{v-v}, \dot{l^*_{v-v}}\}$), removing cell geometry refers to removing ($\{a,\dot{a},p,\dot{p}\}$), and removing dynamics refers to moving ($\{\dot{a},\dot{p}\}$ and $\{\dot{l^*_{v-v}}\}$).

\textit{Dataset.}---
We use a publicly available \textit{Drosophila} gastrulation dataset published in~\cite{stern2022deconstructing}. The raw dataset contains segmented time-lapsed 3-D whole embryo configurations throughout the gastrulation process. Specifically, for each frame, the dataset contains cell coordinates, cell vertex coordinates, cell-cell adjacency, and cell junctions. The data is pre-processed using a customized MATLAB script to arrive at the input format for our geometric deep-learning model. The script for pre-processing the data can be found on our GitHub site.

The synthetic dataset is created using the vertex model following simulation procedures previously described in~\cite{huang2022shear}. In the simulation, a cell monolayer is subjected to external shear strain; at large strain values, an avalanche of local cell rearrangements occurs, which is accompanied by significant drops in tissue stress. Simulations are performed both with fluid-like and solid-like tissues, each at 50 random initial configurations. Extended results on the synthetic dataset can be found in {\color{blue} Supplementary Materials}.

\textit{Experiments.}---
Details of the model architecture and choice of hyperparameters are provided in the {\color{blue} Supplementary Materials}. Unless otherwise noted, all models are trained for 500 epochs on an NVIDIA T4 or L4 GPU.

\textit{Performance.}---
We use the area under the curve (AUC) of the receiver operating characteristic curve and binary cross entropy (BCE) loss to evaluate the model performance.

\textit{Statistics.}---
All statistics are reported as mean $\pm$ std. One-way ANOVA tests are performed when applicable.

\subsection*{Acknowledgements}
We thank L. Yang, W. Lu, and C. Cupo for helpful discussions. This work is supported by NIH (1R01GM140108). M.G. acknowledges the Sloan Research Fellowship. MJB acknowledges support from NIH (5R01AR077793), USDA (2021-69012-35978) and ARO (W911NF2220213). D.B. and A. N. acknowledge support from the National Science Foundation ( DMR-2046683 and PHY-2019745), the Alfred P. Sloan Foundation, the Human Frontier Science Program (Ref.-No.: RGP0007/2022), the National Institutes of Health (R35GM15049), and the Northeastern University Discovery Cluster. We thank Junxiang Huang for providing SPV simulation data.

\subsection*{Code Availability}
Our code will be publicly available on our GitHub repository at:

\url{https://github.com/GuoLab-CellMechanics/Multicell-Fold}.

Trained model weights will be available on Hugging Face.

\bibliographystyle{unsrtnat}
\bibliography{bib}

\end{document}


\title{Supplementary Materials for\\ Multicell-Fold: geometric learning in folding multicellular life}

\author{Haiqian Yang}
\affiliation{Department of Mechanical Engineering, Massachusetts Institute of Technology, 77 Massachusetts Ave., Cambridge, MA 02139, USA}

\author{Anh Q. Nguyen}
\affiliation{Department of Physics, Northeastern University, Boston, MA 02115, USA}

\author{Dapeng Bi}
\affiliation{Department of Physics, Northeastern University, Boston, MA 02115, USA}

\author{Markus J. Buehler}
\affiliation{Department of Mechanical Engineering, Massachusetts Institute of Technology, 77 Massachusetts Ave., Cambridge, MA 02139, USA}
\affiliation{Laboratory for Atomistic and Molecular Mechanics (LAMM), Massachusetts Institute of Technology, 77 Massachusetts Ave., Cambridge, MA 02139, USA}
\affiliation{Center for Computational Science and Engineering, Schwarzman College of Computing, Massachusetts Institute of Technology, 77 Massachusetts Ave., Cambridge, MA 02139, USA}

\author{Ming Guo}\thanks{guom@mit.edu}
\affiliation{Department of Mechanical Engineering, Massachusetts Institute of Technology, 77 Massachusetts Ave., Cambridge, MA 02139, USA}

\date{\today}

\maketitle

\tableofcontents

\appendix
\onecolumngrid
\renewcommand\thefigure{S\arabic{figure}}
\renewcommand{\thetable}{S\arabic{table}}
\setcounter{figure}{0}
\setcounter{table}{0}

\renewcommand{\theequation}{S\arabic{equation}}
\setcounter{equation}{0}



\clearpage
\subsection{Data structure and embedding}
\subsubsection{The dual-graph data structure of multicellular collectives}
We represent the dual graph consisting of cells, vertices, cell-cell adjacencies, and cell junctions as
\begin{equation}
\begin{aligned}
    &\text{Cell positions} \quad \{\mathcal{C}\}\\
    &\text{Vertices positions} \quad \{\mathcal{V}\}\\
    &\text{Cell-cell adjacency} \quad \{\mathcal{E}_{c-c}\}\\
    &\text{Cell junctions} \quad \{\mathcal{E}_{v-v}\}\\
    &\text{Cell-vertice links} \quad \{\mathcal{E}_{c-v}\}
\end{aligned}
\end{equation}
which essentially is a graph $\mathcal{G}$ containing two types of nodes $\{\mathcal{C}\}$ and $\{\mathcal{V}\}$ and three types of edges $\{\mathcal{E}_{c-c}\}$, $\{\mathcal{E}_{v-v}\}$, and $\{\mathcal{E}_{c-v}\}$.

\subsubsection{Structural embedding}
While it is feasible to directly learn from the raw graph $\mathcal{G}$ with raw coordinates, it is beneficial to construct the model inputs from symmetry considerations. Here we focus on invariant geometric quantities.

For the node embedding, we consider
\begin{equation}
\begin{aligned}
    a\\
    p
\end{aligned}
\end{equation}
where $a$ is cell area and $p$ is cell perimeter.

For any type of edge ($\{\mathcal{E}_{c-c}\}$, $\{\mathcal{E}_{v-v}\}$, and $\{\mathcal{E}_{c-v}\}$), we consider their length as the edge attribute. Recent studies have shown that short edges are reminiscent of ``defects" in amorphous materials, which is especially important in cell rearrangement~\cite{nguyen2024origin}, therefore, we use the exponentially weighted edge length as input to the neural network
\begin{equation}\label{eq:SIedgeAttr}
    e^{-l}
\end{equation}
which encourages the model to pay more attention to the patterns of the short edges.

By default, we do not provide edge orientation angle $\theta$, because the angle is not an invariant quantity. In the ablation study below (Fig.~\ref{fig:simu}c\&d), where edge orientation is provided in a model as a comparison, it is important to realize that each edge is non-directional (i.e. edge with orientation $\theta$ and $\theta+\pi$ should be considered as the same); therefore, we consider the tensor product of the unit direction $\mathbf{n}$ weighted by edge length $\quad e^{-l} \mathbf{n} \otimes \mathbf{n}$. Its two independent components are assigned as edge attributes to replace Eq.~\ref{eq:SIedgeAttr}, which are

\begin{equation}
\begin{aligned}
    & e^{-l} \cos{2\theta}\\
    & e^{-l} \sin{2\theta}
\end{aligned}
\end{equation}

In the case of ``full" representation, where multiple types of nodes or edges are used as input, we add one additional node/edge embedding to distinguish the node/edge types
\begin{equation}
\begin{aligned}
    & 1 \quad \text{if} \quad \mathcal{N} \in \{\mathcal{C}\}\\
    & 2 \quad \text{if} \quad \mathcal{N} \in \{\mathcal{V}\}\\
    & 1 \quad \text{if} \quad \mathcal{E} \in \{\mathcal{E}_{c-c}\}\\
    & 2 \quad \text{if} \quad \mathcal{E} \in \{\mathcal{E}_{v-v}\}\\
    & 3 \quad \text{if} \quad \mathcal{E} \in \{\mathcal{E}_{c-v}\}\\
\end{aligned}
\end{equation}

In the case that some features are only defined on one type of node/edge, the rest are padded as zero.

In the ablation study provided in Fig.~\ref{fig:simu}c\&d, the mechanical tension in the simulation is defined as
\begin{equation}
    p_i + p_j - 2 p_0
\end{equation}
where $p_i$ and $p_j$ are the perimeters of the two adjacent cells sharing the junction, and $p_0$ is the target perimeter, which is a coefficient in the free energy of the simulation.

In the \textit{Drosophila} dataset, we have further included the ``general displacement" of all geometric quantities
\begin{equation}
\begin{aligned}
    & \dot{a} = a_t - a_{t-1}\\
    & \dot{p} = p_t - p_{t-1}\\
    & \dot{l^*_{v-v}} = (e^{-l_{v-v}})_t - (e^{-l_{v-v}})_{t-1}
\end{aligned}
\end{equation}
These general displacements are not provided in training the models with synthetic data, because the simulations are performed at quasistatic conditions, that is, between adjacent frames the system is allowed to relax for a significant amount of time.

\subsubsection{Criteria for identifying the loss of cell junctions}
The loss of cell-cell junctions $\{R\}_t$ is computed as the difference between $\{\mathcal{E}_{c-c}\}_t$ and $\{\mathcal{E}_{c-c}\}_{t+1}$,
\begin{equation}
    \{R\}_t = \text{diff} \, (\{\mathcal{E}_{c-c}\}_t, \{\mathcal{E}_{c-c}\}_{t+1}),
\end{equation}
meaning that a pair of cells lose the shared cell-cell junction if it appears as a neighboring pair in frame $t$ but not in frame $t+1$.

\subsection{Synthetic dataset}
The synthetic dataset is generated using the Self-Propelled Voronoi (SPV) model with simulation details in~\cite{huang2022shear}. Previous studies have shown that simulated cell monolayers exhibit distinct physical properties governed by the target shape index (SI); with $\mathrm{SI}<3.81$ the monolayer is solid-like, while with $\mathrm{SI}>3.81$ the monolayer is fluid-like~\cite{huang2022shear,bi2016motility,bi2015density}. Therefore, here we perform simulations of solid-like monolayers with $\mathrm{SI}=3.72$ and fluid-like monolayers with $\mathrm{SI}=3.92$ under quasi-static conditions. Under each condition, the simulation was performed at 50 random seed numbers each with 3000 frames. The simulation contains 400 cells under periodic boundary conditions subjected to external shear strain at step size 0.002 (a representative system configuration is shown in Fig.~\ref{fig:simu}a). Between consecutive strain steps, the system is allowed to relax and reach a steady state. At each strain step, the system configuration and corresponding shear stress $\sigma$ are recorded.

\subsubsection{Identifying efficient representation}

\begin{figure}[h!]
    \centering
    \includegraphics[width=.95\textwidth]{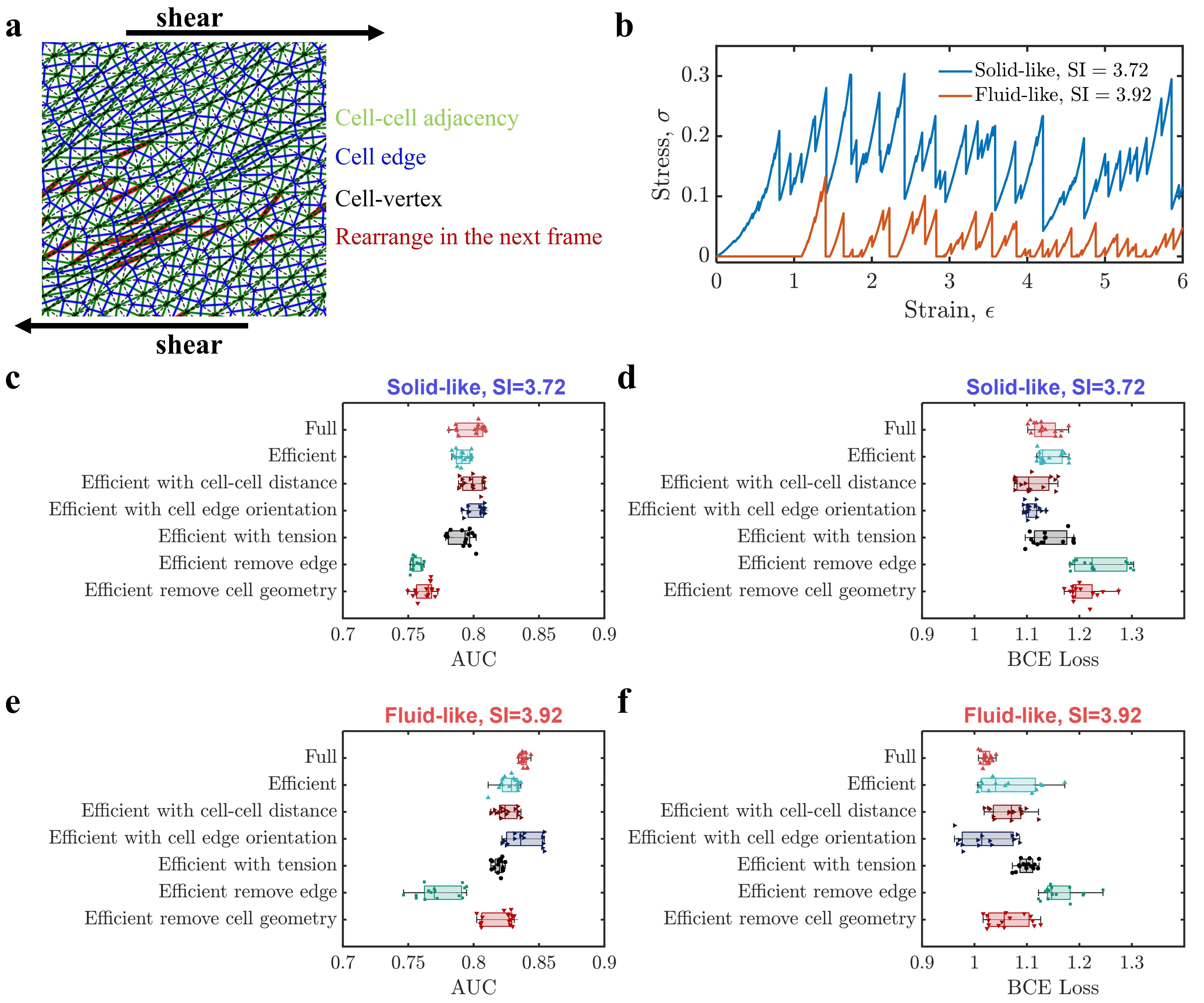}
    \caption{Identifying efficient graph representation in SPV simulations. (a) A representative snapshot of the cell monolayer demonstrating the data structure. As described in detail in the method section, the input data contains cells, vertices, cell-cell adjacency, cell edge, and cell-vertex connection, and the output is a binary label whether the cell-pair will undergo rearrangement in the next frame. (b) Stress-strain relation of a solid-like system ($\mathrm{SI}=3.72$), and a fluid-like system ($\mathrm{SI}=3.92$). (c) AUC and (d) BCE loss of the solid-like system with different graph representations. (e) AUC and (f) BCE loss of the fluid-like system with different graph representations.}
    \label{fig:simu}
\end{figure}

Using the full dual-graph representation, our model is capable of predicting local cell rearrangements prior to their occurrence (Fig.~\ref{fig:simu}c\&d). However, the computation cost of the full representation is very high, therefore, we seek to identify a minimal set of the input dual graphs that can reduce the cost but still with competitive performance. 

Realizing that the cell-edge graph and cell-cell adjacency graph are mathematical duals, our first step is to use cells and cell-cell adjacency as the primary graph, and concatenate the cell-edge information instead on the cell-cell adjacency. Doing so shrinks to 1/3 of its original size. Secondly, we remove each one of the input features and only keep the ones that have a significant impact on the model performance. By doing so, we have identified an ``efficient" representation, with cell area and perimeter as node embedding, and actual cell edge length as edge embedding assigned to the cell-cell adjacency (the actual cell edges are removed from the graph). We show that this efficient representation is slightly worse but almost the same as the full representation, regardless of whether the monolayer is fluid-like or solid-like (Fig.~\ref{fig:simu}c\&d). From this efficient representation, we show that removing either cell-edge length or cell geometry significantly damages the model performance (Fig.~\ref{fig:simu}c\&d). It is worth noting that removing either cell-edge length or cell geometry from the efficient representation is essentially only keeping geometrical information from one of the dual graphs; this highlights the necessity of taking into account the information from both parts of the dual graph.

Adding cell-cell distance or tension back to the input does not improve the model performance (Fig.~\ref{fig:simu}c\&d). While in the literature it has been hypothesized that tension might govern the cell rearrangement, surprisingly, here we do not observe any benefits by directly providing tension as an edge attribute. However, because the free energy in the SPV simulation is a quadratic function of the cell perimeter, the tension in the simulation linearly depends on the perimeter of two adjacent cells; it is probably reasonable to expect that GNN is capable of learning tension within the first convolution. Tension inference tasks and further informing models on cell rearrangement tasks with inferred tension with experimental systems would be interesting for future study. Besides, we indeed observe including edge orientation can improve the performance; this matches our intuition because all the simulations have the same shear boundary condition, in the same direction.

\begin{figure}[h!]
    \centering
    \includegraphics[width=.9\textwidth]{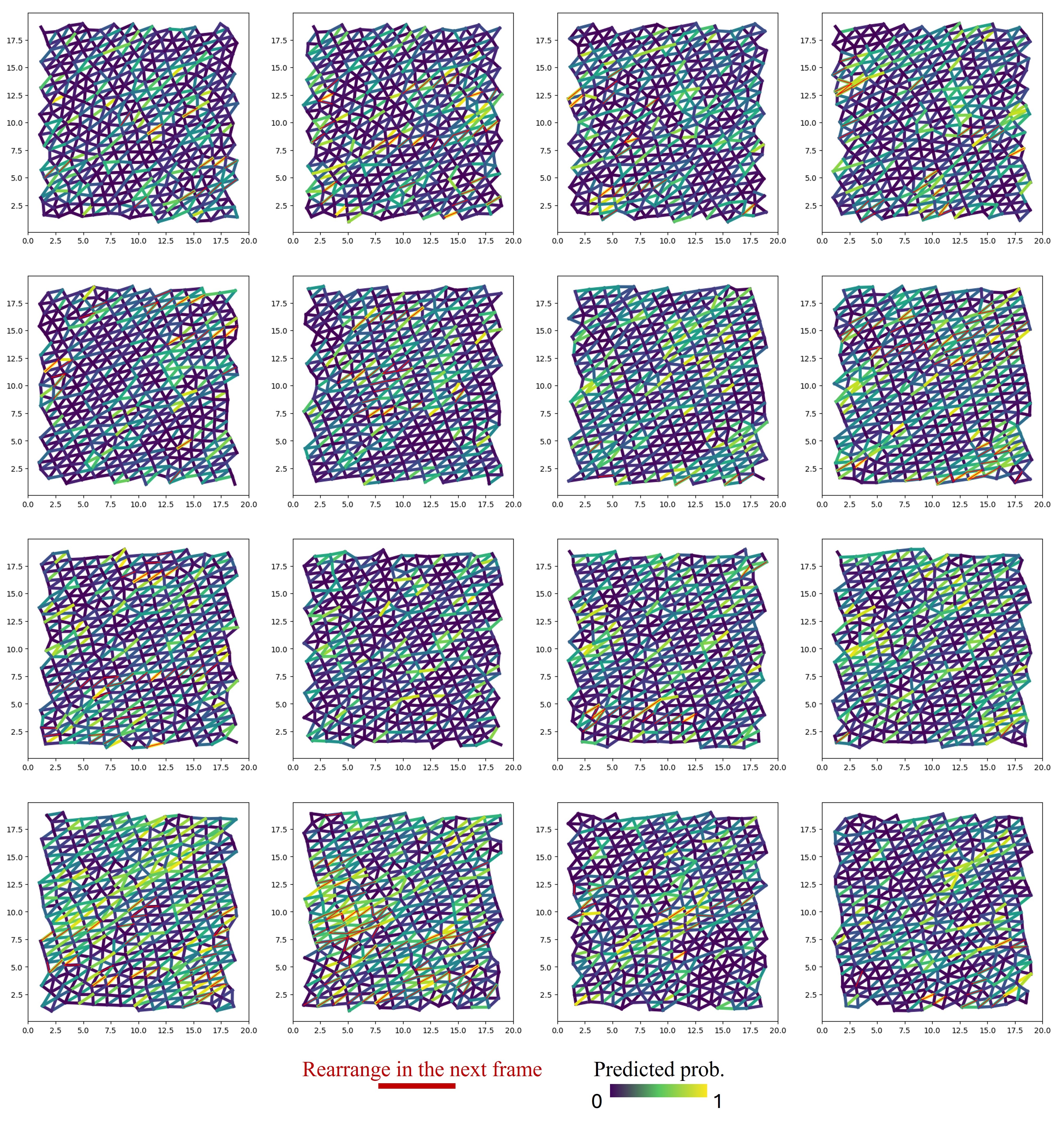}
    \caption{Example predictions. $\mathrm{SI}=3.92$.}
    \label{fig:SIsimuDemo}
\end{figure}

\subsubsection{Predicting tissue-level stresses in the synthetic dataset}
\begin{figure}[h!]
    \centering
    \includegraphics[width=.95\textwidth]{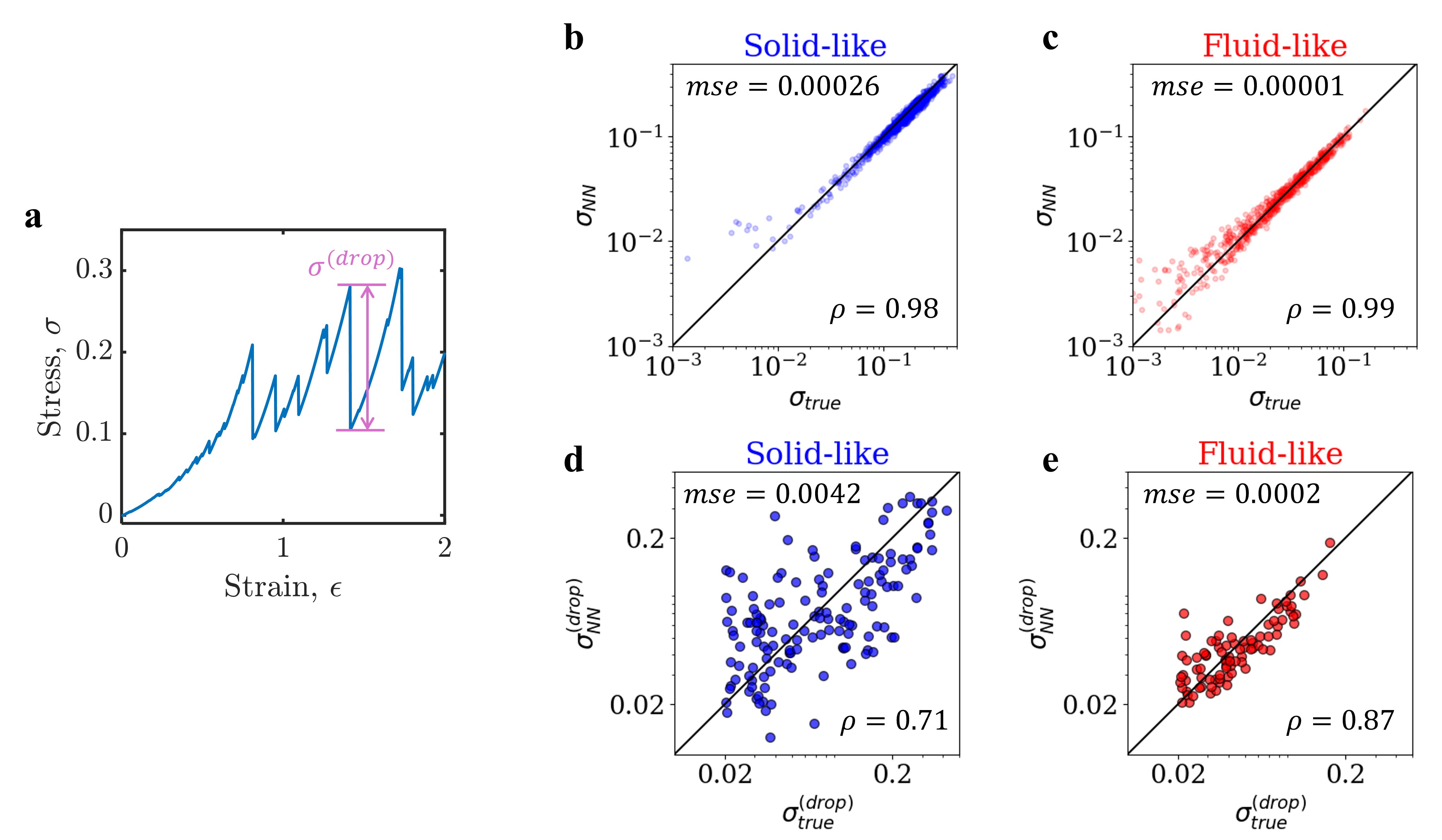}
    \caption{Predicting tissue-level stress. (a) A typical stress-strain relation with the purple highlights the magnitude of the stress drop $\sigma^{(drop)}$. Neural network prediction $\sigma_{NN}$ vs. ground truth $\sigma_{true}$ of same frame stress, in solid-like (b) and fluid-like (c) simulated cell monolayers. Neural network prediction $\sigma^{(drop)}_{NN}$ vs. ground truth $\sigma^{(drop)}_{true}$ of the next-frame stress drop, in solid-like (d) and fluid-like (e) simulated cell monolayers. Solid-like, $\mathrm{SI}=3.72$; fluid-like, $\mathrm{SI}=3.92$. The dropout rate of the graph transformer layer is set to 0 in this case; other hyperparameters are the same.}
    \label{fig:stressdrop}
\end{figure}

Mechanical stresses build up as the monolayer deforms. In theory, the tissue stress can be calculated by taking derivatives of the free energy function with respect to the monolayer configuration. Here, without providing the information of any constitutive relation, we train a model to calculate the stress from the monolayer configuration represented using the efficient graphs identified above. The model achieves highly accurate calculation with a Pearson correlation of 0.98 for the solid-like system and 0.99 for the fluid-like system (Fig.~\ref{fig:stressdrop}).

As the stress increases, the monolayer suddenly fails at some point, and the tissue stress suddenly drops, which is an ``avalanche" type of behavior, accompanied by a large number of local cell rearrangements. Right before the stress drops, the monolayers are at critical configurations. Even though it is quite difficult to analytically predict the magnitude of the stress drop before it happens, intuitively these critical configurations must share some hidden similarities, which could allow predicting how large the stress drop will be. To test this idea, we select those frames with considerable stress drops ($\sigma^{(drop)}>0.02$) and seek to extract such features by training a model to predict the stress drop $\sigma^{(drop)} = \sigma(t)-\sigma(t+1)$. We find that our learning algorithm can achieve reasonable performance with a Pearson correlation of 0.71 for the solid-like system and 0.87 for the fluid-like system (Fig.~\ref{fig:stressdrop}).

\clearpage
\subsection{Extended information and results on the \textit{Drosophila} dataset}
\subsubsection{Extended information of the \textit{Drosophila} dataset}

\begin{figure}[h!]
    \centering
    \includegraphics[width=.95\textwidth]{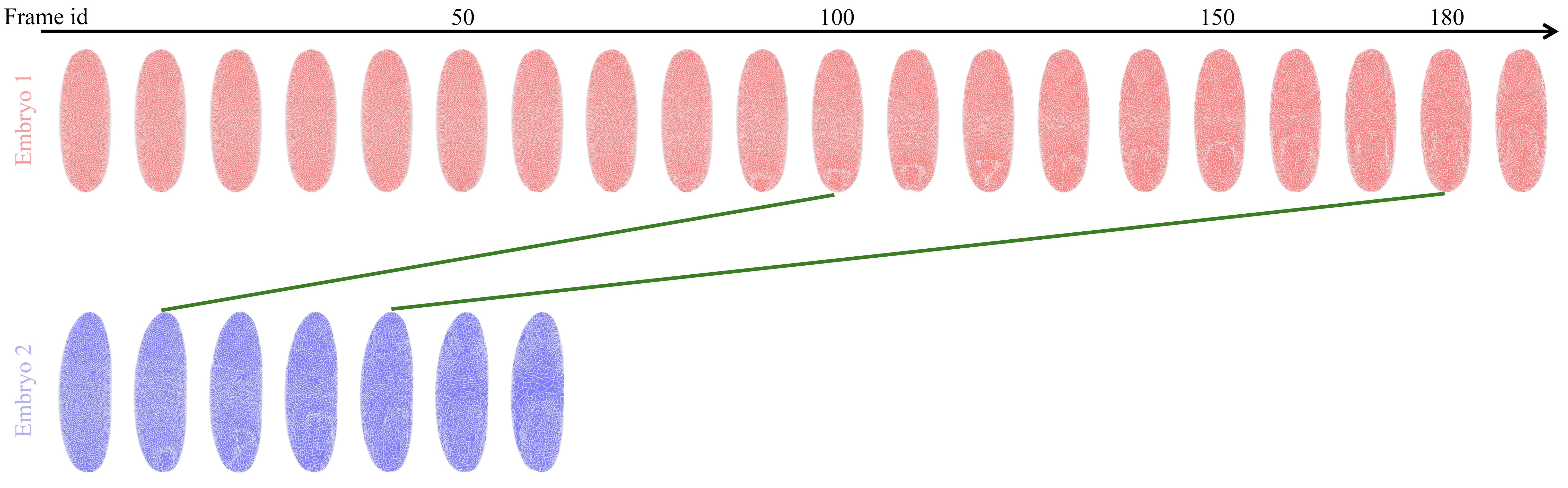}
    \caption{Example frames of the two embryo sequences. The two green lines highlight the frame pairs with similar morphology by manually examining the two sequences.}
    \label{fig:DrosophilaSequence}
\end{figure}

We use the dataset published in~\cite{stern2022deconstructing}, which contains segmented time-lapsed 3-D whole-embryo configuration of two \textit{Drosophila} embryos. Example snapshots of the two embryos are shown in Fig.~\ref{fig:DrosophilaSequence}. Note that the two embryos were imaged at different frame rates and angles, and they started their development at different times relative to the starting time of recording.

This dataset is preprocessed to match the input format of our workflow, and the statistics of the two embryos are shown in Table.~\ref{tab:Table1}.

\begin{table}[h]
    \centering
    \begin{tabular}{|c|c|c|c|c|c|}
        \hline
                  &Source    & number of frames &  total nodes & total edges    & total rearrangement events detected\\
        \hline
        Train     &Embryo\#1 &       191        &  943045      &   2820573      & 126849        \\ 
        Test      &Embryo\#2 &        59        &  252557      &    754528      &  41544        \\
        \hline
    \end{tabular}
    \caption{Statistics of the \textit{Drosophila} embryo dataset.}
    \label{tab:Table1}
\end{table}

\clearpage
\subsubsection{Additional results}

\begin{table}[h]
    \centering
    \begin{tabular}{c|c|c}
         \hline
         AUC                & BCE               & Accuracy  \\
         \hline
         $0.913 \pm 0.004$  & $0.809 \pm 0.026$ & 82.67\% (rearrange) and 83.37\% (non-rearrange)\\
         \hline
    \end{tabular}
    \caption{Performance on local cell rearrangements in a developing \textit{Drosophila} embryo.}
    \label{tab:MetricsSummary}
\end{table}

\begin{figure}[h!]
    \centering
    \includegraphics[width=.6\textwidth]{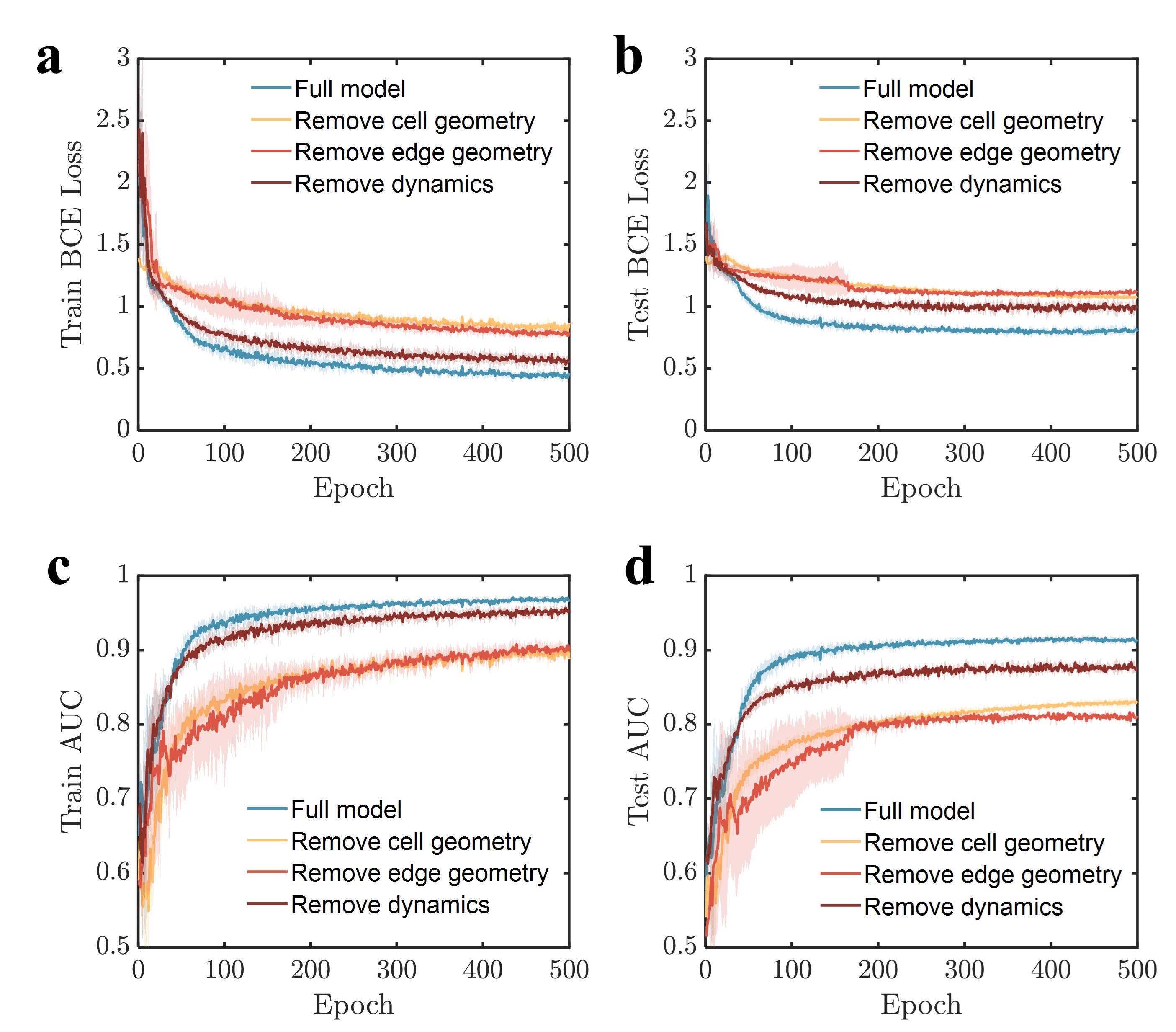}
    \caption{Performance over epochs. BCE loss of the train set (a) and test set (b). AUC of the train set (c) and test set (d). In Fig.~{\color{blue}4} in the main text we compare the last 5 epochs of the model performance.}
    \label{fig:DrosophilaEpochs}
\end{figure}

\begin{figure}[h!]
    \centering
    \includegraphics[width=.7\textwidth]{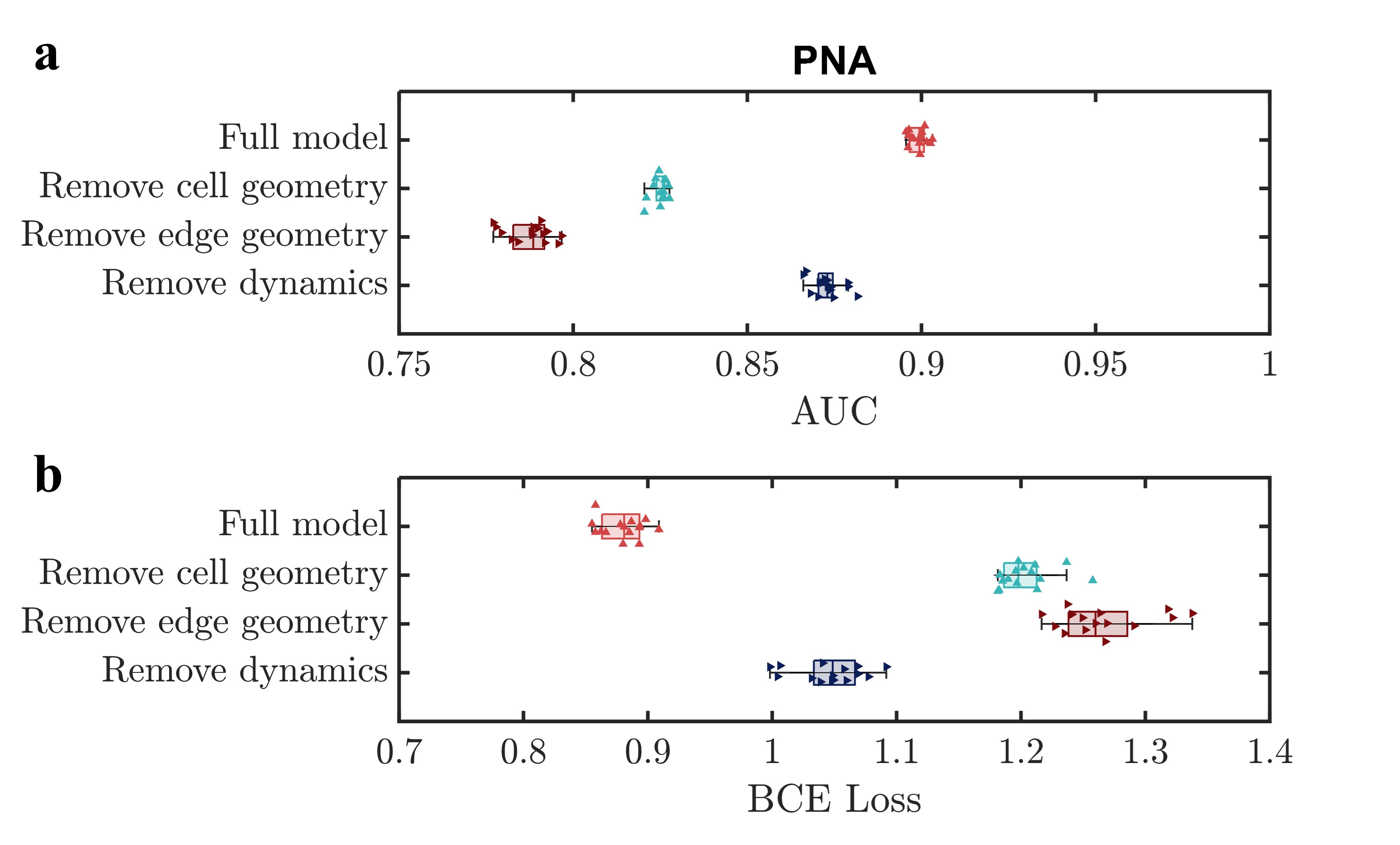}
    \caption{Ablation test on the \textit{Drosophila} dataset with PNA convolutional layers instead of graph transformer layers. (a) AUC performance. (b) BCE loss.}
    \label{fig:DrosophilaPNA}
\end{figure}

\begin{figure}[h!]
    \centering
    \includegraphics[width=.7\textwidth]{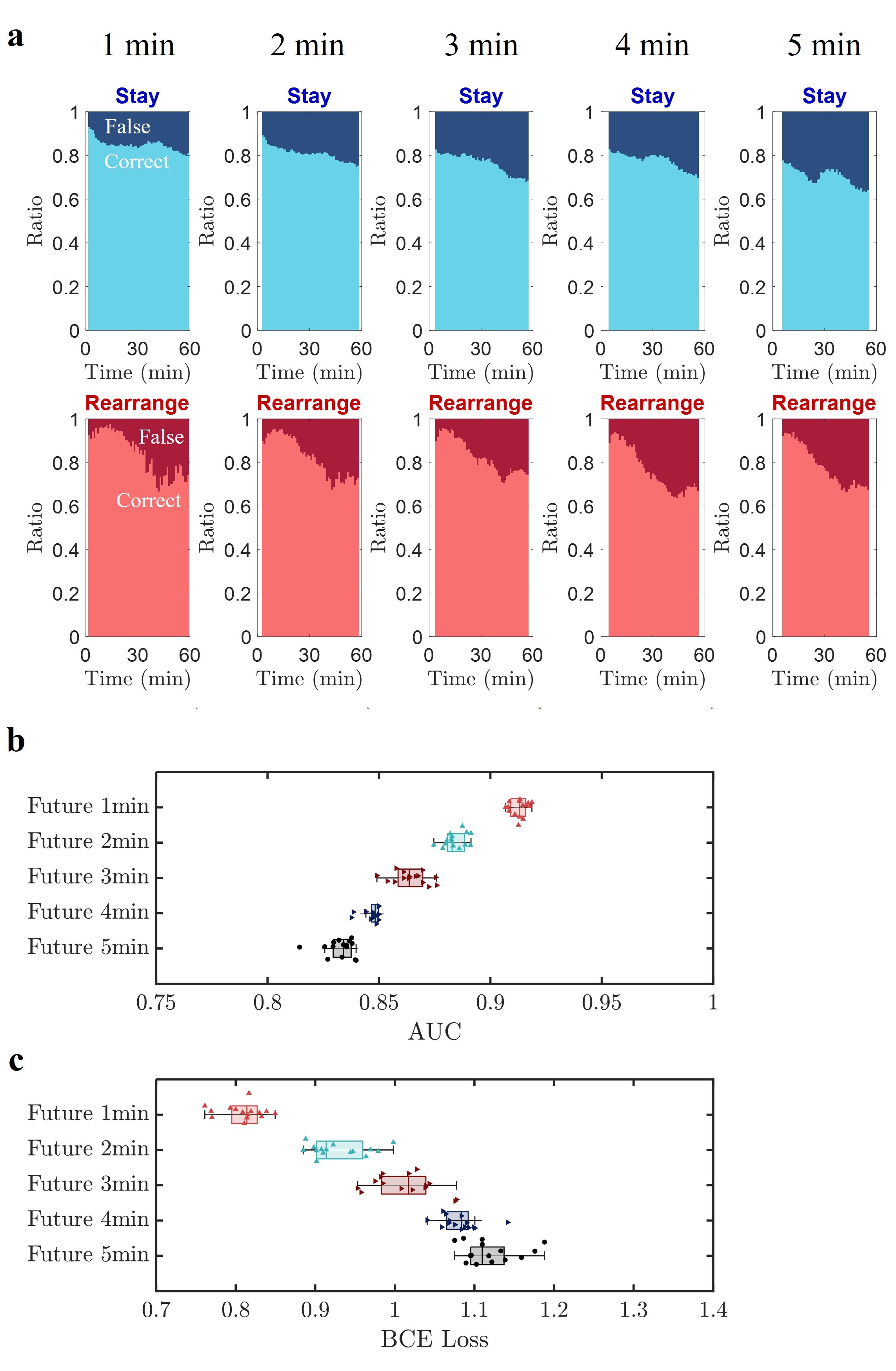}
    \caption{Predicting further into the future. (a) Accuracy at different future time intervals. (b) AUC and (c) BCE loss at different future time intervals.}
    \label{fig:DrosophilaLongerFuture}
\end{figure}

\clearpage
\subsection{Details of the model architecture and influence of hyperparameters}
\subsubsection{Model architecture for edge predictions}
\begin{figure}[h!]
    \centering
    \includegraphics[width=.95\textwidth]{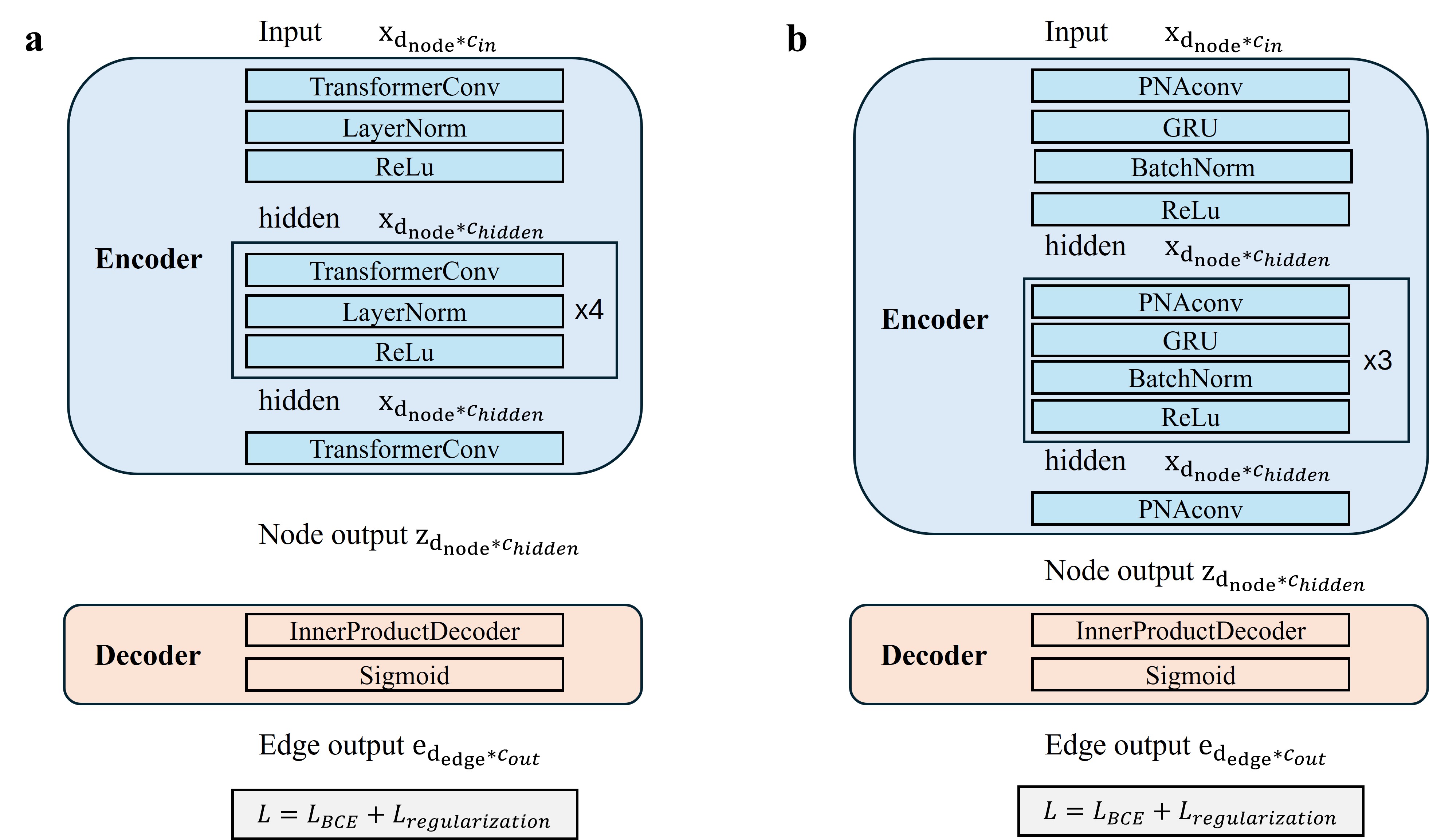}
    \caption{Detailed model architecture for edge predictions. (a) Using Graph Transformer convolutional layers. (b) Using PNA convolutional layers.}
    \label{fig:SImodel}
\end{figure}

A detailed model structure is shown in Fig.~\ref{fig:SImodel}.

\textit{Encoder.}---
We use either graph transformer~\citep{shi2020masked} or Principal Neighbourhood Aggregation (PNA)~\citep{corso2020principal} convolutional layers in the encoder block. Unless otherwise noted, for the tasks on the \textit{Drosophila} dataset, the PNA models consist of 5 PNA convolutional layers and 15 channels, and the graph transformer models consist of 6 graph-transformer convolutional layers, 24 channels, and 4 attention heads with 0.5 drop-out rate; on the synthetic dataset, the graph transformer models consist of 5 graph-transformer convolutional layers with other hyperparameters the same. In all cases, GRU layers are used between PNA layers as residual connections; batch norm layers are applied between PNA layers and layer norm layers are applied between graph-transformer layers.

\textit{Decoder.}---
For edge prediction, we employ a graph autoencoder structure~\cite{kipf2016variational}. Briefly, the output of the encoder is then fed into an inner product decoder. We sum over the hidden channels and take the sigmoid function of the output of the decoder to generate edge prediction.

\textit{Classification.}---
For calculating the percentage accuracy, we classify the cell pairs with predicted rearranging probability $>0.5$ as the predicted rearranging group, and the rest as the static group.

\textit{Visualization.}---
In Fig.~\ref{fig:SIsimuDemo}, the predicted rearranging probability is directly visualized on cell-cell adjacency and compared to the rearranging edges.
In Fig.~{\color{blue}3} in the main text, we further visualize the rearranging probability on cells instead. In this case, a cell is marked as predicted rearranging with high confidence if it with any of its neighbors is predicted with a higher than 0.85 probability of rearranging.

\subsubsection{Model architecture for tissue-level predictions}
We start by employing the same graph encoder as described above. On the output of the graph encoder, we apply global mean pooling, followed by a linear layer to get a single-valued tissue-level prediction. 

To visualize the activation map, we denote the output of the encoder as $\mathbf{z}^{(i)}$, where the superscript $^{(i)}$ indicates the \textit{i}-th channel, and each $\mathbf{z}^{(i)}$ is a vector containing the activation on each node; the weights of the linear layer are denoted as $w_l^{(i)}$. The activation map is calculated as $\mathbf{A} = \sum_{i} w_l^{(i)} \mathbf{z}^{(i)}$. The activation map is normalized as $\mathbf{A}^*=\mathbf{A}/\max(\mathbf{A})$. $\mathbf{A}^*$ is then used to color code the cell polygons on the original graphs in Fig.~{\color{blue}2} in the main text.

\subsubsection{Experiments}

\textit{Drosophila tasks.}---
On the experimental dataset, we use embryo\#1 for training and embryo\#2 for testing. All cell pairs are included during training and testing without any negative sampling.

\textit{Synthetic tasks.}---
On the synthetic dataset, the simulations are performed at 50 random seed numbers; we use seeds 1-45 for training, and 46-50 for testing. Among the 3000 frames for each seed, we sample only those when stress drops in the next frame (with a threshold of 0.005). Among all the cell pairs, on this synthetic dataset the positive group (rearranging pairs) only comprises a small portion; to deal with this data imbalance during edge prediction, we employ negative sampling, that is, we only sample the equal amount of negative cell pairs (not rearranging) as the positive cell pairs (rearranging) during training and validation.

\textit{Training details.}---
Mean squared error is used for the regression tasks, and binary cross entropy loss is used for predicting local cell rearrangements. AdamW~\cite{loshchilov2017decoupled} is used with learning rate 5e-4 and weight decay 1e-8. All models are trained for 500 epochs on a NVDIA T4 or L4 GPU.

\subsubsection{Regularization and hyperparameters}

In training the model for predicting cell rearrangements, we use the total loss
\begin{equation}
\begin{aligned}
    &L_{total} = L_{BCE} + L_{regularization}\\
    &\text{with} \quad L_{regularization} = \left( \bar{e} - \frac{N^{(e)}_{pos}}{N^{(e)}_{tot}} \right)^2
\end{aligned}
\end{equation}
where $\bar{e}$ is the average of the predicted probability of cell rearrangement over the whole batch, $N^{(e)}_{pos}$ is the number of rearranged cell pairs and $N^{(e)}_{tot}$ is the total number of cell pairs. The regularization loss is included to encourage the model to prioritize predicting the right number of cell rearrangements.

\begin{figure}[h!]
    \centering
    \includegraphics[width=.5\textwidth]{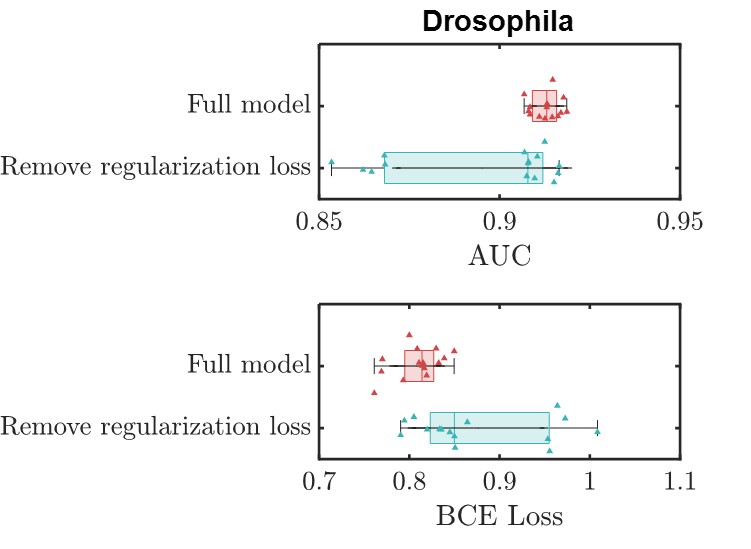}
    \caption{The influence of the regularization loss.}
    \label{fig:SIloss2}
\end{figure}

In Table~\ref{tab:TableS1}, we report the model performance using different hyperparameters.

\begin{table}[h]
    \centering
    \begin{tabular}{c|cccccccc}
    \hline
    Type of change   &layer & channel & dropout & head & AUC & AUC std & BCE & BCE std \\ \hline 
      -         &6 & 24 & 0.5 & 4 & 0.917 & 0.002 & 0.779 & 0.017 \\ 
    layer       &3 & 24 & 0.5 & 4 & 0.794 & 0.005 & 1.361 & 0.024 \\ 
    layer       &4 & 24 & 0.5 & 4 & 0.869 & 0.001 & 0.943 & 0.002 \\ 
    layer       &5 & 24 & 0.5 & 4 & 0.910 & 0.003 & 0.800 & 0.007 \\ 
    layer       &7 & 24 & 0.5 & 4 & 0.908 & 0.002 & 0.904 & 0.023 \\ 
    layer       &8 & 24 & 0.5 & 4 & 0.870 & 0.003 & 1.065 & 0.032 \\ 
    channel     &6 & 16 & 0.5 & 4 & 0.909 & 0.004 & 0.821 & 0.025 \\ 
    channel     &6 & 32 & 0.5 & 4 & 0.901 & 0.002 & 0.928 & 0.010 \\ 
    channel     &6 & 8 & 0.5 & 4 & 0.901 & 0.002 & 0.850 & 0.009 \\ 
    heads       &6 & 24 & 0.5 & 2 & 0.771 & 0.006 & 1.205 & 0.023 \\ 
    heads       &6 & 24 & 0.5 & 6 & 0.910 & 0.004 & 0.898 & 0.032 \\ 
    dropout     &6 & 24 & 0.0 & 4 & 0.804 & 0.001 & 2.983 & 0.044 \\ 
    dropout     &6 & 24 & 0.3 & 4 & 0.890 & 0.002 & 0.922 & 0.010 \\ 
    dropout     &6 & 24 & 0.6 & 4 & 0.906 & 0.002 & 0.837 & 0.013 \\ 
    dropout     &6 & 24 & 0.7 & 4 & 0.849 & 0.012 & 1.081 & 0.037 \\ 
    dropout     &6 & 24 & 0.8 & 4 & 0.885 & 0.002 & 0.922 & 0.012 \\ 
    \hline
    \end{tabular}
    \caption{Effect of hyperparameters. The models are trained for 500 epochs using random seed number 42 and statistics of the last 5 epochs are reported here.}
    \label{tab:TableS1}
\end{table}

\clearpage

\subsection{Caption of Supplementary videos}
\textbf{Supplementary video 1.} Predicting local cell rearrangement 1 minute in the future, in a 3-D \textit{Drosophila} embryo over time. The predicted rearranging cells are marked in orange, the predicted non-rearranging cells are marked in blue. Black circles mark the ground-truth rearranging cells within the next 1 minute.

\clearpage

\bibliographystyle{unsrtnat}
\bibliography{bib}